\documentclass[aps,prb,twocolumn,superscriptaddress,showkeys]{revtex4-1}

\usepackage{amsmath,amssymb,graphicx,upgreek}

\begin{document}

\title{Slow-light plasmonic metamaterial based on dressed-state analog of electromagnetically-induced transparency}

\author{S{\o}ren Raza}
\email{Corresponding author: sraz@iti.sdu.dk}
\author{Sergey I. Bozhevolnyi}
\affiliation{Department of Technology and Innovation, University of Southern Denmark, Niels Bohrs All\'{e} 1, DK-5230 Odense M, Denmark}

\begin{abstract}
We consider a simple configuration for realizing one-dimensional slow-light metamaterials with large bandwidth-delay products using stub-shaped Fabry-Perot resonators as building blocks. Each metaatom gives rise to large group indices due to a classical analog of the dressed-state picture of electromagnetically-induced transparency. By connecting up to eight metaatoms, we find bandwidth-delay products over unity and group indices approaching 100. Our approach is quite general and can be applied to any type of Fabry-Perot resonators and tuned to different operating wavelengths.
\end{abstract}

\maketitle
Electromagnetically-induced transparency (EIT) occurs in atomic systems due to the destructive quantum interference of coupled atomic transitions~\cite{Boller:1991,Harris:2008}. The EIT phenomenon renders an otherwise optically opaque medium transparent in a narrow spectral region, which, due to the Kramers--Kronig relations, is accompanied by strong material dispersion, resulting in a significant reduction of the speed of light~\cite{Hau:1999} or even stopping light~\cite{Longdell:2005,Heinze:2013}. For all-optical processing systems, controlling the speed of light is an essential ability, especially for designing optical storage systems, such as optical buffers~\cite{Khurgin:2005}. Unfortunately, atomic EIT often requires extreme environmental conditions, such as cryogenic temperatures, which are not suitable for chip-based systems. In contrast, classical analogs of EIT in metal-based plasmonic structures~\cite{Zhang:2008,Papasimakis:2008,Liu:2009a,Kekatpure:2010,Bozhevolnyi:2011,Han:2011} pave the way for exploiting the reduction of speed of light on the nanoscale. Such extreme control of light is possible due to the coupling of light to surface plasmons propagating at metal-dielectric interfaces, thereby squeezing the light beyond the diffraction limit~\cite{Gramotnev:2010}. 

In the quantum mechanical description of atomic EIT, two theoretical interpretations based on either the bare states or the dressed states are available. While both are equivalent in the quantum description, their classical analog differ significantly. In particular, the classical EIT analog of the bare states interpretation requires the use of radiant and subradiant plasmonic modes~\cite{Zhang:2008,Papasimakis:2008,Liu:2009a,Liu:2009b}, while the dressed states picture is based on detuned radiant resonators~\cite{Bozhevolnyi:2011,Han:2011,Chen:2012}. The dressed-states analog has recently received substantial attention, especially by utilizing detuned resonators coupled to metal-insulator-metal (MIM) waveguides~\cite{Han:2011,Huang:2011,Chen:2012b,Lu:2012,Zhu:2014,Cao:2014}. The gap surface plasmon mode supported by the MIM waveguide offers a good compromise between mode confinement and propagation length, making it an ideal candidate for slow-light devices. 

One of the parameters for characterizing a slow-light device is the minimal group velocity $v_\textrm{g}$, or, equivalently, the largest group index $n_\textrm{g}=c/v_\textrm{g}$, which can be attained. However, a large group index is only practical if available in a wide frequency bandwidth $\Delta f$. Hence, an important figure-of-merit for slow-light applications is the bandwidth-delay product $\tau \Delta f$, where $\tau = v_\textrm{g} l$ is the delay experienced by an optical pulse over a propagation length $l$~\cite{Baba:2008}. To ensure the delay of a pulse of duration $\sim \Delta f^{-1}$, we must have $\tau \Delta f \gtrsim 1$. In this work, we present a simple configuration for generating over-unity bandwidth-delay products by creating one-dimensional metamaterials using two detuned Fabry-Perot resonators as our metaatom. We study detuned stub resonators aperture-coupled to a MIM waveguide to realize the dressed-state analog of atomic EIT with accompanying large group index. By joining up to eight metaatoms, we show the expected convergence of both the real and imaginary parts of the metamaterial effective refractive index. We find a group index approaching 100 and bandwidth-delay product of 1.2 with a reasonable transmittance of $10\%$.
\begin{figure}[t]
	\centering
	\includegraphics[scale=1]{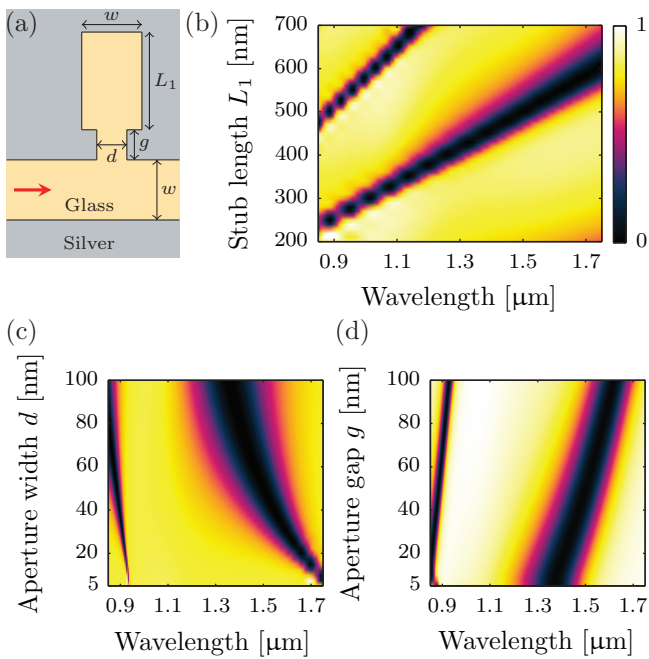}
	\caption{(a) Schematic illustration of silver-glass-silver waveguide aperture-coupled (with width $d$ and gap $g$) to a single stub-shaped Fabry-Perot resonator of length $L_1$. Transmittance through waveguide (color plot) as a function of wavelength and (b) stub length $L_1$, (c) aperture width $d$, and (d) aperture gap $g$. The following parameters are used, unless they are varied in the plot: $w=100$~nm, $d=g=50$~nm, and $L_1=500$~nm.}
	\label{fig:fig1}
\end{figure}

We consider a silver-glass-silver waveguide, where the refractive index of glass is 1.45, while the dielectric function for silver is interpolated from tabulated values~\cite{Johnson:1972}. The width of the waveguide is fixed to $w=100$~nm throughout this study. Using a port excitation scheme in a commercial finite-element frequency-domain solver (COMSOL Multiphysics 5.0), we launch the gap surface plasmon mode of the MIM waveguide and detect the transmittance at an output port positioned after the stub resonators. Before discussing the system consisting of two detuned stub resonators, we first optimize the case of a single stub resonator of length $L_1$, which is coupled through an aperture of width $d$ and gap $g$ to the MIM waveguide, as illustrated in Fig.~\ref{fig:fig1}(a). In Fig.~\ref{fig:fig1}(b) we vary the length of the stub for fixed aperture dimensions of $g=d=50$~nm. For $L_1<500$~nm, we notice a single dip in the transmittance spectra due to the excitation of an anti-symmetric Fabry-Perot mode supported by the stub, which exhibits a standing wave pattern with two antinodes in the magnetic field (not depicted). The transmittance dip is due to the destructive interference of the bus MIM wave and the reflected wave from the stub resonator. For $L_1>500$~nm an additional transmittance dip at shorter wavelength shows up since the stub resonator now also supports a symmetric Fabry-Perot mode with three antinodes. For increasing stub length, both Fabry-Perot modes redshift and broaden. By varying the stub length, we can tune the resonance of a given Fabry-Perot mode to a desired wavelength. In the subsequent analysis, we choose $L_1=500$~nm since both modes are then present in the considered wavelength range. In Fig.~\ref{fig:fig1}(c) and (d) we study the effect of the aperture width $d$ and gap $g$, respectively. As an important feature, we see that decreasing the aperture width provides a narrower transmittance dip since the reflection from the stub surface next to the aperture increases. Put in other words, the finesse of the stub-shaped Fabry-Perot resonator increases with decreasing aperture width. As shown in earlier studies~\cite{Han:2011}, a spectrally narrow transmittance dip is beneficial since it produces a strong EIT effect with resulting large group index. However, we find that for too small values of $d$ the transmittance minima begin to increase (i.e., the transmittance does not reach zero at the resonance wavelengths) due to the decrease in coupling between the MIM waveguide and the stub resonator. Hence, a trade-off between the transmittance minimum and the spectral width of the transmittance dip (i.e., quality factor) exists, which can be controlled through the aperture width. Finally, as shown in Fig.~\ref{fig:fig1}(d) increasing the aperture gap (for a fixed aperture width) only produces a redshift of the transmittance dip (and no significant change in the spectral width) as a consequence of additional phase accumulation in the aperture, which slightly modifies the resonance condition of the aperture-coupled stub system. Based on this analysis, we keep the same value for the aperture gap (i.e., $g=50$~nm) but decrease the width to $d=30$~nm for a narrow transmittance dip which still reaches zero at the Fabry-Perot resonance wavelengths.

\begin{figure}[t]
	\centering
	\includegraphics[scale=1]{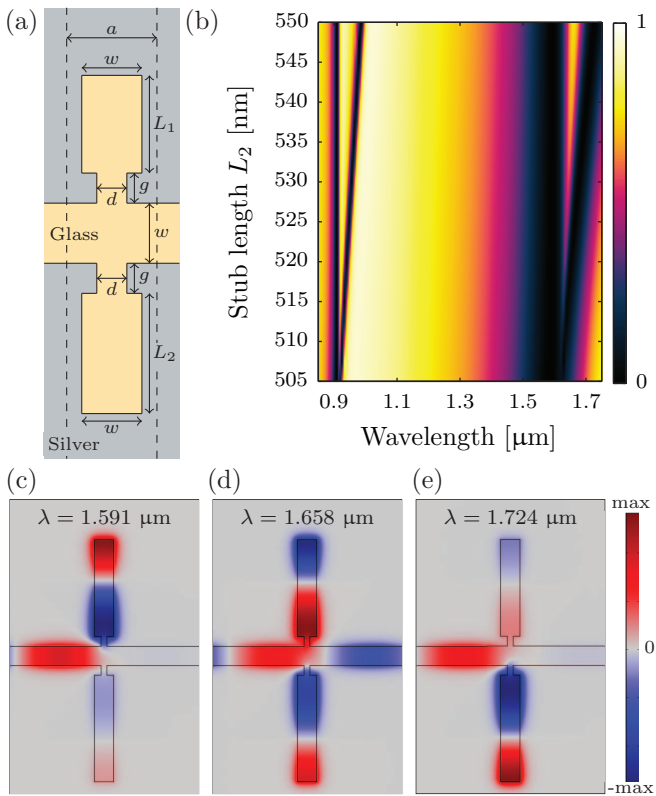}
	\caption{(a) Schematic illustration of silver-glass-silver waveguide aperture-coupled (with width $d$ and gap $g$) to two stub-shaped Fabry-Perot resonators of lengths $L_1$ and $L_2$. The metaatom cell size $a$ is also shown. (b) Transmittance through the waveguide (color plot) as a function of wavelength and second stub length $L_2$. The fixed parameters are: $w=100$~nm, $d=30$~nm, $g=50$~nm, and $L_1=500$~nm. (c-e) Out-of-plane magnetic field component at the resonance wavelength of the short stub, the wavelength of maximum transmittance in the EIT-like window, and the resonance wavelength of the long stub, respectively. The length of the second stub is $L_2=550$~nm.}
	\label{fig:fig2}
\end{figure}

Next, we add an additional stub of length $L_2$ on the opposite side of the waveguide, as illustrated in Fig.~\ref{fig:fig2}(a). The two apertures, which couple to each of the stub resonators, have the same dimensions. We will later consider the double-stub system as our metaatom (with unit size $a$) when building our one-dimensional slow-light metamaterial. For now, we study a single cell. To generate the EIT effect based on the dressed state interpretation we need two detuned resonators~\cite{Han:2011,Bozhevolnyi:2011}, which can be achieved by simply varying the length of the second stub resonator, while keeping the widths of the two stubs identical (the opposite approach will also give rise to detuned resonators). Figure~\ref{fig:fig2}(b) shows transmittance spectra of the double-stub system for $L_2$ values which are slightly larger than $L_1$ ($L_1 = 500$~nm). As the difference between the lengths of the two stubs $L_2-L_1$ increases, the detuning of the two resonators increases, creating an EIT window in the spectral regions of the transmittance dips. We note that the long-wavelength transmittance dip requires a larger detuning to create the EIT window, since the dip is spectrally wider. In fact, to have a sufficiently large transmittance maxima in the EIT window ($\gtrsim80\%$) we find that $L_2$ should be at least 550~nm. In Fig.~\ref{fig:fig2}(c-e) we show the magnetic field profile of the EIT phenomenon for three different wavelengths for the case of $L_2=550$~nm. For wavelengths shorter [Fig.~\ref{fig:fig2}(c)] and longer [Fig.~\ref{fig:fig2}(e)] than the EIT window, the two stub resonators are excited independently, producing in both cases a weak transmission through the waveguide. At the transmittance maximum in the EIT window [Fig.~\ref{fig:fig2}(d)], both stub resonators are excited simultaneously but with opposite phase, thereby cancelling the destructive interference normally created by the individual stubs. A similar picture is seen for the low-wavelength EIT window (with three antinodes in the magnetic field profile of each stub instead of two), showing that the EIT effect created due to the detuned resonators is independent of the stub mode.

\begin{figure}[t]
	\centering
	\includegraphics[scale=1]{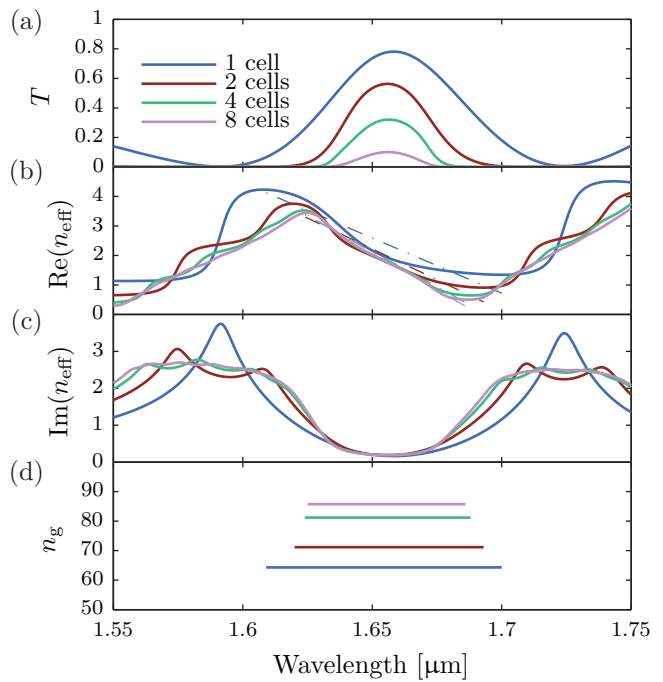}
	\caption{(a) Transmittance $T$ through one-dimensional slow-light metamaterial consisting of one to eight double-stub metaatoms. (b-c) Real and imaginary parts of the effective index $n_\textrm{eff}$ of the metamaterial, respectively. In the spectral regions of normal dispersion, a linear fit of the real part of the effective index is performed (dashed lines). (d) Group index $n_g$ determined from the linear fits of the real part of the effective index. The metaatom consisting of two detuned aperture-coupled stubs is sketched in Fig.~\ref{fig:fig2}(a) with the following parameters: $w=100$~nm, $d=30$~nm, $g=50$~nm, $L_1=500$~nm, $L_2=550$~nm, and $a=200$~nm.}
	\label{fig:fig3}
\end{figure}

In Fig.~\ref{fig:fig3} we focus on the long-wavelength EIT window [Fig.~\ref{fig:fig3}(a)] and determine the effective refractive index $n_\textrm{eff}$ [Fig.~\ref{fig:fig3}(b-c)] along with the group index $n_\textrm{g}$ [Fig.~\ref{fig:fig3}(d)]. The real part of the effective index is calculated from the phase difference of the magnetic fields $\Delta\phi$ at positions before and after the metaatoms as $\textrm{Re}(n_\textrm{eff}) =  \Delta\phi / (N a k_0)$, where $N$ is the number of metaatoms and $k_0=2\pi c/\lambda$ is the vacuum wave vector~\cite{Han:2011,Santillan:2011,Santillan:2014}. The imaginary part of the effective index is determined from the transmittance $T$ as $\textrm{Im}(n_\textrm{eff}) = -\ln(T)/(2Nak_0)$, assuming an exponential decay of the power of the fields. Finally, the group index is given as $n_\textrm{g} = \textrm{Re}(n_\textrm{eff}) - \lambda \tfrac{ \textrm d [ \textrm{Re}(n_\textrm{eff})]}{\textrm d \lambda}$ in the spectral regions of normal dispersion $\tfrac{ \textrm d [ \textrm{Re}(n_\textrm{eff})]}{\textrm d \lambda}<0$. To avoid unphysical large variations in the group index, we perform a linear fit of the real part of the effective index in the regions of normal dispersion [shown as dashed lines in Fig.~\ref{fig:fig3}(b)]. Figure~\ref{fig:fig3} shows that a single double-stub metaatom produces a group index larger than 60 in a bandwidth of approximately 50~nm [determined as the full-width at half-maximum (FWHM) of the transmittance peak], with a transmittance maximum close to $80\%$.

The metamaterial description is only applicable for metaatoms significantly smaller than the operating wavelength, i.e., $a\ll\lambda$, setting an upper limit to our cell size. Simultaneously, too small cell sizes give rise to undesired coupling between the metaatoms. Despite these two limits the cell size is still quite tunable, and we find that a cell size of $a=200$~nm provides a good trade-off. Figure~\ref{fig:fig2}(b-c) shows that the effective index quickly converges with increasing number of cells, supporting the metamaterial interpretation. Additionally, the group index [Fig.~\ref{fig:fig3}(d)] also increases with the number of cells, reaching close to a two-order decrease in the speed of light with only 8 cells. However, due to the increase in propagation length $l=Na$ the FWHM and maximum transmittance decrease. As discussed earlier, an important figure-of-merit for any slow-light device is the bandwidth-delay product, which we estimate using the relation $\tau \Delta f = l \Delta\lambda [ n_\textrm{g}(\lambda_\textrm{T}) - n_\textrm{MIM}(\lambda_\textrm{T}) ] / \lambda_\textrm{T}^2$. Here, $\Delta\lambda$ is the FWHM of the transmittance peak, while the group indices of the metamaterial $n_\textrm{g}$ and the MIM waveguide $n_\textrm{MIM}$ are evaluated at the wavelength of maximum transmittance $\lambda_\textrm{T}$. For a single metaatom we find the bandwidth-delay product to be 0.26, which is quite large considering the short propagation length. Joining additional metaatoms significantly increases the bandwidth-delay product to 0.36, 0.68, and 1.2 for 2, 4, and 8 cells, respectively. Despite the decrease in FWHM of the transmittance peak, the increase in propagation length and in group index results in an overall increase in the bandwidth-delay product. The large values for the bandwidth-delay product come at the expense of decreasing maximum transmittance. However, even for 8 cells with a huge bandwidth-delay product of 1.2, the maximum transmittance is still $10\%$. Adding more cells increases the bandwidth-delay product even further, but the transmittance level becomes impractical. Given a certain tolerance level for the maximum transmittance, say $10\%$, the maximum number of joined metaatoms can actually be estimated from the imaginary part of the effective index of a \emph{single} metaatom [Fig.~\ref{fig:fig3}(c)] as $N = -\ln(T)/[2\textrm{Im}(n_\textrm{eff}) a k_0 ]$. This is possible due to the fast convergence of the effective index in the EIT window with increasing number of cells. 

While we only showed the long-wavelength EIT window in Fig.~\ref{fig:fig3}, we have also performed simulations on the short-wavelength EIT window of Fig.~\ref{fig:fig2}. Using the same approach for creating the metamaterial, we find comparable values for the group index and bandwidth-delay product, providing support that our slow-light metamaterial can be made to operate using different Fabry-Perot modes at different wavelengths. In fact, the detuned resonators of the individual metaatoms can be achieved with other Fabry-Perot resonators than the aperture-coupled stub-shaped considered here, such as nanodisk cavities~\cite{Lu:2012} or T-shaped resonators~\cite{Wen:2014}.

In summary, we have considered a simple configuration for designing slow-light metamaterials with large bandwidth-delay products based on metaatoms exhibiting EIT. The EIT effect is achieved using detuned resonators to mimic the atomic dressed-state picture. By adjusting the detuning, the metamaterial can be tuned to different operating wavelengths.

S.~I.~B. acknowledges financial support by European Research Council, Grant 341054 (PLAQNAP).

\end{document}